\documentclass[final,numberedheadings]{aipproc}

\usepackage{ }

\layoutstyle{6x9}

\newcommand{\hide}[1]{}
\renewcommand{\texttt}[1]{{\tt\small{#1}}}

\begin{document}

\title{The Mock LISA Data Challenges: An Overview}



\author{The \emph{Mock LISA Data Challenge Task Force}: Keith A. Arnaud}{address={Gravitational Astrophysics Laboratory, NASA Goddard Space Flight Center, \\ 8800 Greenbelt Rd., Greenbelt, MD 20771, USA}}
\author{Stanislav Babak}{address={Max-Planck-Institut f\"ur Gravitationsphysik (Albert-Einstein-Institut), \\ Am M\"uhlenberg 1, D-14476 Golm bei Potsdam, Germany}}
\author{John G. Baker}{address={Gravitational Astrophysics Laboratory, NASA Goddard Space Flight Center, \\ 8800 Greenbelt Rd., Greenbelt, MD 20771, USA}}
\author{Matthew J. Benacquista}{address={Center for Gravitational Wave Astronomy, University of Texas at Brownsville, \\ Brownsville, TX 78520, USA}}
\author{Neil J. Cornish}{address={Department of Physics, Montana State University, Bozeman, MT 59717, USA}}
\author{Curt Cutler}{address={Jet Propulsion Laboratory, California Institute of Technology, Pasadena, CA 91109, USA}}
\author{Shane L. Larson}{address={Center for Gravitational Wave Physics, 104 Davey Laboratory, University Park, PA 16802, USA},altaddress={Department of Physics, Weber State University, 2508 University Circle, Ogden, UT 84408, USA}}
\author{B. S. Sathyaprakash}{address={School of Physics and Astronomy, Cardiff University, Cardiff, CF243YB, UK}}
\author{Michele Vallisneri}{address={Jet Propulsion Laboratory, California Institute of Technology, Pasadena, CA 91109, USA}}
\author{Alberto Vecchio}{address={School of Physics and Astronomy, University of Birmingham, \\ Edgbaston, Birmingham B152TT, UK}}
\author{Jean--Yves Vinet}{address={Department ARTEMIS, Observatoire de la C\^ote d'Azur, BP 429, 06304 Nice, France}}

\begin{abstract}
The LISA International Science Team Working Group on Data Analysis (LIST-WG1B) is sponsoring several rounds of mock data challenges, with the purpose of fostering the development of LISA data analysis capabilities, and of demonstrating technical readiness for the maximum science exploitation of the LISA data. The first round of challenge data sets were released at the Sixth LISA Symposium. We briefly describe the objectives, structure, and time-line of this programme.
\end{abstract}

\maketitle

\section{Overview}
\label{s:intro}

The Laser Interferometer Space Antenna (LISA) is a space borne gravitational wave (GW) observatory designed for detailed studies of a wide variety of gravitational wave sources throughout the Universe in the frequency range 0.1 mHz -- 0.1 Hz~\cite{lisa_ppa}. LISA is an all-sky monitor with the capability of directly measuring source parameters, such as masses, spins and distances, with unprecedented precision. The LISA data set is anticipated to contain a large number $(\sim 10^4)$ of resolvable overlapping sources, ranging from galactic sub-solar mass binary systems to high-redshift massive black holes; moreover the mission has the potential of discovering radically new classes of sources, such as GW primordial stochastic backgrounds, cosmic strings and exotic compact objects~\cite{CT2002}. Most sources detectable by LISA are long lived compared to the mission lifetime ($>$ 2 yr) and the data will contain strong GW foregrounds generated by abundant populations of galactic and extra-galactic white-dwarf binary systems and possibly  solar-mass compact objects captured by massive black holes in galactic nuclei. Indeed there is no established expertise for this kind of data set, although much experience is gained in the analysis of GW data collected by ground-based detectors (see Saulson's contribution in this volume). In ground-based observations GWs are rare and weak, whereas in the low frequency band they are numerous and (a fair fraction of them) yield high signal-to-noise ratios (SNR). It is therefore vital in preparation for the mission to tackle these new analysis problems and develop the necessary methodologies for the maximum science exploitation of such a revolutionary data set.

At the LISA International Science Team (LIST) meeting of December 2005, the Working Group on Data Analysis (LIST-WG1B) decided to embark in the organisation of several rounds of mock data challenges (MLDC), with the dual purpose of (i)  fostering the development of LISA data analysis tools and capabilities, and (ii) demonstrating the technical readiness already achieved by the gravitational-wave community in distilling a rich science payoff from the LISA data output. The LISA Mock Data Challenges were also proposed and discussed at meetings organized by the US and European LISA Project that were attended by a broad cross section of the international GW community. These challenges are meant to be blind tests, but not really contests; the greatest scientific benefit stemming from them will come from the quantitative comparison of results, analysis methods, and implementations.

A Mock LISA Data Challenge Task Force was constituted at the beginning of 2006 and has been working since then, to formulate challenge problems of maximum efficacy, to establish criteria for the evaluation of the analyses, to develop standard models of the LISA mission and GW sources, to provide computing tools -- LISA response simulators, source waveform generators, and a Mock Data Challenge file format -- and more generally to provide any technical support necessary to the challengers. The challenges involve the distribution of several data sets, encoded in a simple standard format, and containing combinations of realistic simulated LISA noise with the signals from one or more GW sources of parameters unknown to the challenge participants, who are asked to return the maximum amount of correct information about the sources, and to produce technical notes detailing their work. In this short contribution we summarise the objectives, structure, and time-line of the Mock LISA Data Challenges. In the companion contribution of the Task Force in this volume~\cite{mldcproc2} we provide more technically orientated details of the MLDCs.  Details can be found on the official MLDC website \cite{mldcweb}, in the living \emph{Omnibus} document for Challenge-1 \cite{omnibus}, and on the MLDC Task Force wiki \cite{mldcwiki}.

\section{Structure and timeline}

The MLDCs consist in detecting GW signals embedded in the distributed (mock) LISA data sets and then extracting the maximum amount of information about the source(s) that generate those signals. In order to ensure the incremental development of data analysis approaches, software, pipelines and infrastructure, the Task Force has decided to issue several rounds of challenges, at intervals of approximately 6 months, that contain progressively more complex GW signals and/or noise realisations. Each round of MLDCs consists of multiple data sets with signals of different nature and strength embedded in synthetic noise. Two classes of data sets are distributed at each release: the proper challenge data sets (where the source parameters are unknown) and training data sets, with GW signals of similar nature to those included in the blind tests but whose parameters are made public. All the software and the necessary documentation to generate data sets are public~\cite{mldcweb} so that interested parties can produce their own mock data streams; this should facilitate the development of analysis tools and the testing and validation of the algorithms. The challenge data sets are generated by two members of the Task Force who do {\it not} take part to the challenges and are the only repositories of the key to recover the source parameters. Progress on the work is monitored through extended teleconferences. MLDC results, including analyses of how well different methods performed, will be disseminated through technical documents and articles in peer reviewed journals. Prospective challenge participants are asked to subscribe to the \texttt{lisatools-challenge} mailing list (see~\cite{mldcweb}), which offers a natural communication link between the Task Force and the broader community.

In the first round of challenges (Challenge-1), the data sets contain a single signal or a small number of non-overlapping multiple signals (with one important exception) embedded in Gaussian and stationary noise with no contribution from galactic and/or extragalactic foregrounds. Of course, these challenges are significantly easier than the ``real thing,'' but the goal is to foster the development and validation of basic tools and  building blocks for LISA data analysis, as well to give participants who are new to LISA a chance to ``get up to speed.'' The first-round data sets were realised in June 2006 at the end of the Sixth LISA Symposium and results are due on December 1, 2006. They will be presented to the broad community and discussed in a dedicated session at the 11th Gravitational-Wave Data Analysis Workshop (December 2006). 

The next set of challenges (Challenge-2), expected to be released in December 2006 with results due in June 2007, is regarded by the Task Force as a key milestone in this effort: the goal is to tackle a \emph{global} analysis problem -- which is the distinctive feature of the LISA mission from a GW data analysis point of view -- in the presence of foregrounds within a restricted parameter space in order to limit CPU intensive jobs at this early stage. We expect the results of Challenge-2 to provide a proof of concept of the analysis approach for LISA which can serve as a basis for more detailed studies and further development. Although a final decision on the exact GW content of this data set has yet to be made -- progress on Challenge-1 will surely affect the decision on the format of Challenge-2 -- the data stream will contain a stochastic foreground (produced by a galactic population synthesis model), LISA verification binaries, a small frequency band with strongly overlapping yet resolvable stellar-mass binary systems, a few coalescences from massive black hole binary systems and $\sim 1$ extreme-mass ratio inspiral (EMRI). Moreover the total number of sources in the data set will be unknown. Other data sets containing just one source -- such as EMRIs, stochastic backgrounds, short broad-band bursts -- will also be produced to tackle source specific analysis problems.

The third round of MLDCs (Challenge-3), expected to be released in June 2007 with results due in December 2007, will focus on testing an even more realistic data analysis scenario, introducing non-Gaussianity/non-stationarity in the data set, data gaps, more general gravitational waveforms and a larger number of sources; the details will clearly be influenced by the results of Challenge-1/2. MLDCs will continue beyond the third round, but the exact schedule and format has not been discussed yet.

\section{The first round of challenges}

The first round of challenges is aimed at enabling the development of the necessary data analysis building blocks for simple LISA data analysis tasks. It focuses on data sets that contain a single source (generally with moderate-to-high SNR) or a small number of sources that do not overlap in the frequency domain; an important exception is however made for two data sets that contain a few tens of galactic binary systems whose signals overlap significantly in a small frequency region. These latter data sets provide a fairly realistic representation of the novel aspect in GW data analysis offered by the LISA mission, though restricted to a limited frequency band and a particular signal class. It should therefore provide an early assessment of the soundness of techniques that have been investigated so far to tackle this particular problem. The first round of challenges concentrates on sources currently listed as minimum science requirements~ \cite{sciencereq}: galactic binaries, including verification binaries, and massive black hole binary systems. Data sets containing EMRIs are also made available now, in order to facilitate the development of data analysis tools for this particular demanding class of sources. However, the results of the challenge are due only in June 2007 (the due date for Challenge-2).

All the data sets, with the exception of the EMRI challenge data discussed above, are 1-yr long and contain instrumental noise modelled as Gaussian and stationary with the assumption that the laser frequency noise has been exactly removed. No foreground radiation is included. More details are provided in the companion contribution in this volume~\cite{mldcproc2} and in Refs.~\cite{mldcweb,omnibus}.

\subsection{Galactic stellar mass binaries}

Seven distinct data sets are released containing radiation from stellar-mass galactic binary systems. For Challenge-1 we decided to approximate the waveform as exactly monochromatic in the source reference frame: the signal is therefore described by seven parameters~\cite{mldcproc2}. The first three data sets each contain the signal from a single galactic binary (with optimal SNR for a single TDI output $\approx 20$ to facilitate the analysis) whose geometrical parameters are drawn randomly over the whole relevant rage. The three signals differ primarily in the frequency, which was chosen randomly in the interval 0.9 mHz -- 1.1 mHz, 2.9 mHz -- 3.1 mHz, and 9 mHz -- 11 mHz, respectively. 

A data set was released containing ``verification binaries'': for the MLDCs we define verification binaries as stellar mass binary systems whose orbital period and location in the sky are \emph{exactly} known, whereas the other parameters are totally unknown -- this is a fairly realistic approximation of the actual known systems in the Galaxy, although depending on the specific source some information on other parameters are available and can be incorporated into the analysis. We injected 6 signals from real verification binaries (whose supposedly unknown parameters were randomly drawn from uniform distributions over the relevant range) -- RXJ0806, V407Vul, ESCet, AMCVn , HPLib, and EIPsc -- and 14 synthetic verification binaries that were selected randomly using a binary-population-synthesis code. In all cases we ensured that the coherent SNR over 1 yr of integration was $> 10$. In order to incrementally increase the difficulty of the analysis process, we also released a data set containing 20 unknown galactic binaries whose parameters were all chosen randomly and with frequency in the range 0.1 mHz -- 10 mHz. 

The next two (and final) data sets for galactic binaries included in Challenge-1 are of somewhat different nature and offer a realisation of the main new analysis challenge of the mission: an unknown number of overlapping sources in time and frequency space. We generated and released 2 data streams each containing in a small portion of frequency space a number of sources comparable to what predicted by our present understanding of galactic sub-stellar mass binary populations. The challenge participants are only given the range of the number of binaries present in the data stream (between 40 and 60) and the frequency band in which they are located: between 3 mHz and 3.015 mHz (for the mildly overlapping sources) and 2.9985 mHz and 3.0015 mHz for the strongly overlapping sources. All the other parameters are chosen randomly, with the only constraint being that the coherent SNR using a single TDI data stream exceeds 5. 

\subsection{Massive black hole binary systems}

Two data sets containing a single MBHB inspiral were produced for the first round of challenges. The waveforms are approximated at the restricted second post-Newtonian order (see~\cite{mldcproc2,omnibus,mldcweb} for more details) with no spin-orbit nor spin-spin induced modulations. Each waveform is described by 9 parameters; the masses were chosen having uniform distribution in the range 1--5 $\times 10^6\,M_\odot$ for the primary object and a mass ratio in the range 1--4. The main difference between the two data sets concern the time at which coalescence takes place and the optimal SNR: the first data set was chosen to contain about half a year of effective inspiral (with coalescence time unknown within a range of 40 days) and a signal-to-noise ratio $\approx 500$; the signal injected in the second data set contains an earlier stage of the life time of a binary (with coalescence occurring about $35 \pm 20$ days after the end of the data taking) and a reduced signal-to-noise ratio (in the range $\approx 20-100$). 

\subsection{Extreme mass ratio inspirals}

Finally, with Challenge-1 we are also releasing data sets containing signals from EMRIs. At present, this source class is not included in the minimum science requirements. However the scientific payoff from such observations is so high, and the analysis problem so demanding, that the Task Force felt it was important to release early-on data sets that could help focus the development of EMRI data analysis techniques. We have generated five 2-yr long data sets containing EMRIs with SNR in the range $20 - 120$. The theoretical computation of the EMRI waveforms represents in itself a major challenge which needs to be met before the LISA launch; at present we do not have waveforms that can be used for information extraction. We therefore decided to adopt ``analytical kludge waveforms''~\cite{barack}, which are fastest to compute, while retaining most of the qualitative features of the actual signals (for more details see~\cite{mldcproc2,omnibus,mldcweb}). The signals depend on 14 parameters (we assume the captured compact object to have negligible spin), which were chosen in an astrophysically realistic range: the total and reduced mass of the system were drawn from a uniform distribution in the range $9.5\times 10^5 - 1.05 \times 10^6\,M_\odot$ and $9.5 - 10.5 \,M_\odot$, respectively; the spin of the central massive object was chosen in the range $0.5 - 0.7 \,M^2$ and the orbital eccentricity at plunge to be in the range $0.1 - 0.25$. The plunge was selected to take place between 1.5 yr and 2 yr + 2 weeks from the beginning of the observation. 


\begin{theacknowledgments}
M.V.'s work was supported by the LISA Mission Science Office and by the Human Resources Development Fund at the Jet Propulsion Laboratory, California Institute of Technology, where it was performed under contract with the National Aeronautics and Space Administration.
\end{theacknowledgments}


\begin{thebibliography}{9}
%
\bibitem{lisappa} LISA Study Team, ``LISA: Laser Interferometer Space Antenna for the detection and observation of gravitational waves,'' Pre-Phase A Report, 2nd ed. (Max Planck Institut f\"ur Quantenoptik, Garching, Germany, 1998).
%
\bibitem{CT2002} C. Cutler, and K. S. Thorne, in \emph{General Relativity and Gravitation}, N. T. Bishop and D. M. Sunil, eds. (World Scientific, Singapore, 2002), p. 72.
%
\bibitem{mldcproc2} MLDC Task Force, ``A How-To for the Mock LISA Data Challenges,'' in this volume.
%
\bibitem{mldcweb} Mock LISA Data Challenge Homepage, \url{astrogravs.nasa.gov/docs/mldc}.
%
\bibitem{omnibus} Mock LISA Data Challenge Task Force, ``Document for Challenge 1,'' \url{svn.sourceforge.net/viewvc/lisatools/Docs/challenge1.pdf}.
%
\bibitem{mldcwiki} Mock LISA Data Challenge Task Force wiki, \url{www.tapir.caltech.edu/dokuwiki/listwg1b:home}.
%
\bibitem{sciencereq} T. A. Prince and K. Danzmann, ``LISA Science Requirement Document,'' v. 3.0, 12 May 2005, at \url{www.srl.caltech.edu/lisa/documents}.
%
\bibitem{barack} L. Barack and C. Cutler, \emph{Phys. Rev. D} \textbf{69}, 082005 (2004).

\end{thebibliography}
\end{document}